\documentclass[aps,showpacs,a4paper,floatfix,twocolumn,prc,amsmath,amssymb]{revtex4-1}
\usepackage{amsmath}
\usepackage{graphicx}
\usepackage[lofdepth,lotdepth]{subfig}
\usepackage{ulem}
\usepackage{morefloats}
\usepackage{rotating}
\usepackage{relsize}
\usepackage{color}
\usepackage{floatflt,epsfig}
\usepackage{ulem}
\usepackage{natbib}
\usepackage{filecontents}
\usepackage[utf8]{inputenc}
\usepackage[T1]{fontenc}
\begin{document}

\title{Neutron star cooling and GW170817 constraint within quark-meson coupling models}

\author{Odilon Louren\c{c}o, C\'esar H. Lenzi, Mariana Dutra, Tobias Frederico}
\affiliation{\mbox{Departamento de F\'isica, Instituto Tecnol\'ogico de Aeron\'autica, DCTA, 12228-900, S\~ao Jos\'e dos Campos, SP, Brazil}}

\author{Mrutunjaya Bhuyan}
\affiliation{\mbox{Department of Physics, Faculty of Science, University of Malaya, Kuala Lumpur 50603, Malaysia} \\
\mbox{Institute of Research and Development, Duy Tan University, Da Nang 550000, Vietnam}}

\author{Rodrigo Negreiros}
\affiliation{Instituto de F\'isica, Universidade Federal Fluminense 20420, Niter\'oi, RJ, Brazil}

\author{Cesar V. Flores, Guilherme Grams, D\'ebora P. Menezes}
\affiliation{\mbox{Departamento de F\'isica, Universidade Federal de Santa Catarina, Florian\'opolis, SC, CP 476, CEP 88.040-900, Brazil}}

\begin{abstract}
In the present work we used five different versions of the quark-meson coupling (QMC) model to compute astrophysical quantities related to the GW170817 event and to neutron star cooling process. 
Two of the models are based on the original bag potential structure and three versions consider a harmonic oscillator potential to confine the quarks. The bag-like models also incorporate the pasta phase used to describe the inner crust of neutron stars. We show that the pasta phase always play a minor or negligible role in all studies. Moreover, while no clear correlation between the models that satisfy the GW170817 constraints and the slope of the symmetry energy is found, a clear correlation is observed between the slope and the fact that the cooling is fast or slow, i.e., fast (slow) cooling is related to higher (lower) values of the slope. We did not find one unique model that can describe, at the same time, GW170817 constraints and give a perfect description of the possible cooling processes.
 
\end{abstract}

\maketitle

\section{Introduction}

The observation of the binary neutron star system GW170817 by the LIGO-VIRGO scientific collaboration and also in the X-ray, ultraviolet, optical, infrared, and radio bands gave rise to the new era of multimessenger astronomy. 
All these joint observations gave support to the idea that
GW170817 was produced when two neutron stars merged in NGC4993. The observation of the electromagnetic counterpart of GW170817 by the  Fermi Gamma-ray observatory corroborated that binary neutron star mergers are associated to short gamma ray bursts (sGRB) and kilonova emissions probably powered by the radioactive decay of r-process nuclei synthesized in the ejecta \cite{Abbott_2017}. Therefore multimessenger  observations are an excellent tool to extract precious information of compact stars under extreme conditions. 

In the next years more signal detections will be possible by further upgrades of dubbed A+ and LIGO Voyager \cite{LIGOthecnicalReport} and by the joint collaboration of KAGRA and LIGO India \cite{Aasi2015,Acernese_2014,PhysRevD.88.043007}. In addition, planned third generation observatories such as Einstein Telescope (ET) and Cosmic Explorer (CE) may bring us the detection of binary neutron stars at cosmological distances \cite{,PhysRevD.96.084004,PhysRevD.96.084039,Punturo_2010}. 

As can be seen the observation of binary systems in different channels is of outstanding importance in order to establish stronger constraints on neutron star physics. As a matter of fact neutron stars have been, for many decades,  objects of intense astrophysical research.  Electromagnetic observations were used to establish some limits on the mass and radius of neutron stars. In light of the recent detections, the gravitational wave channel has become a new way to observe NS and also provide new neutron star observables, like the tidal polarizability arising just before the merging.

The dimensionless tidal deformability or tidal polarizability (TP) is related to the induced  deformation that a neutron star undergoes by the influence of the tidal field of its neutron star companion in the binary system \cite{1995MNRAS.275..301K}. That influence is expected to be detected in the low frequency range of the inspiral stage, where the effect  is  a  small  correction  in  the  waveform phase. As each different neutron star composition has a characteristic response to the tidal field, the TP can be used to discriminate between  different equations of state (EOS). Therefore, this new and very interesting physics offers an exciting possibility to investigate the neutron star composition in a very clear form. 

In addition, thermal evolution studies are a complementary way of probing properties of the neutron star. The investigation of the cooling of compact stars has been proved a promising method for exploring the properties of neutron stars, as it is strongly connected to both micro and macroscopic realms  \citep{TSURUTA1965,Maxwell1979,Page2006,Weber2007,Negreiros2010,2018ApJ...863..104N}.  There is a wealth of literature on the cooling of neutron stars and related phenomena such as magnetic field \citep{Aguilera2008,Pons2009,Niebergal2010a,Negreiros2017b}, deconfined quark matter \citep{Horvath1991,Blaschke2000a,Shovkovy2002,Grigorian2005,Alford2005a}, superfluidity \citep{Levenfish1994,Schaab1997,Alford2005,Page2009}, rotation \citep{Negreiros2012,Negreiros2013,Negreiros2017}, among others \citep{Weber2005,Alford2005,Gusakov2005,Negreiros2010}.

In the present work we present five different versions of the quark-meson coupling (QMC) model equations of state chosen to describe neutron star (NS) matter. The first two are respectively the original QMC model with an underlying bag structure and its counterpart with the inclusion of an interaction between the meson $\omega$ and $\rho$ fields. In both cases, the pasta phase is also considered in the description of the NS inner crust. The other three versions are modified QMC (MQMC), where the parameters are adjusted so that the
constituent quarks are confined to a flavor-independent harmonic oscillator potential~\cite{jpg,npa}. 

The above mentioned models are confronted with recent astrophysical constraints, including the ones predicted by the GW170817. We also perform cooling simulations for all models and compare the results with observed data. This study will allow us (in conjunction with the aforementioned studies) a better evaluation of the quality of the underlying microscopic models adopted.

The formalism used for the EOS is shown in section \ref{qmc}, the necessary equations to estimate the quantities related to the GW constraints are given in section \ref{GW} and for the cooling process in section \ref{cooling}. We present and discuss our results in section \ref{results} and in the last section we make our final remarks.

\section{Equations of State}
\label{qmc}

In this section we present the formalism of the original QMC model, its counterpart with the inclusion of the $\omega\rho$ interaction and in the sequel, the modified QMC model. 

\subsection{QMC and QMC$\omega\rho$ models}

In the QMC model, the nucleon in nuclear medium is assumed to be a
static spherical MIT bag in which quarks interact with the scalar ($\sigma$)
and vector ($\omega$, $\rho$) fields, and those
are treated as classical fields in the mean field
approximation (MFA) \cite{guichon}.
The quark field, $\psi_{q_{N}}$, inside the bag then
satisfies the equation of motion:
\begin{eqnarray}
\left[i\,\rlap{/}\partial \right.&-&(m_q-g_\sigma^q\,)-g_\omega^q\, \omega\,\gamma^0\nonumber \\
&+&\left.  \frac{1}{2} g^q_\rho \tau_z \rho_{03}\gamma^0\right]
\,\psi_{q_{N}}(x)=0\ , \quad  q=u,d
\label{eq-motion}
\end{eqnarray}
where $m_q$ is the current quark mass, and $g_\sigma^q$,
$g_\omega^q$ and $g_\rho^q$
denote the quark-meson coupling constants. The normalized ground state for a quark in the bag 
is given by
\begin{eqnarray}
\psi_{q_{N}}({\bf r}, t) &=& {\cal N}_{q_{N}} \exp 
\left(-i\epsilon_{q_{N}} t/R_N \right) \nonumber \\
&\times& \left(
\begin{array}{c}
  j_{0_{N}}\left(x_{q_{N}} r/R_N\right)\\
i\beta_{q_{N}} \vec{\sigma} \cdot \hat r j_{1_{N}}\left(x_{q_{N}} r/R_N\right)
\end{array}\right)
 \frac{\chi_q}{\sqrt{4\pi}} ~,
\end{eqnarray}
where
\begin{equation}
\epsilon_{q_{N}}=\Omega_{q_{N}}+R_N\left(g_\omega^q\, \omega+
 \frac{1}{2} g^q_\rho \tau_z \rho_{03} \right) ,
\end{equation}
and,
\begin{equation}
\beta_{q_{N}}=\sqrt{\frac{\Omega_{q_{N}}-R_N\, m_q^*}{\Omega_{q_{N}}\, +R_N\, m_q^* }}\ ,
\end{equation}
with the normalization factor given by
\begin{equation}
{\cal N}_{q_{N}}^{-2} = 2R_N^3 j_0^2(x_q)\left[\Omega_q(\Omega_q-1)
+ R_N m_q^*/2 \right] \Big/ x_q^2 ~,
\end{equation}
where $\Omega_{q_{N}}\equiv \sqrt{x_{q_{N}}^2+(R_N\, m_q^*)^2}$,
$m_q^*=m_q-g_\sigma^q\, \sigma$, $R_N$ is the
bag radius of nucleon $N$ and $\chi_q$ is the quark spinor. The bag eigenvalue for nucleon $N$, $x_{q_{N}}$, is determined by the
boundary condition at the bag surface
\begin{equation}
j_{0_{N}}(x_{q_{N}})=\beta_{q_{N}}\, j_{1_{N}}(x_{q_{N}})\ .
\label{bun-con}
\end{equation}
%

%%---------------------tabela bulk qmc-----------------------------------
\begin{table*}[htb!]
\caption{Nuclear matter and stellar properties obtained with the QMC and QMC$\omega\rho$
  models. } 
\begin{tabular}{lccccccccccc}
\hline
\hline
Model   &$B/A$ & $n_0$   &  $\Lambda_v$ & $g_\rho$ & $M_N^*/M_N$  & $J$ &  $L_0$   &   $K_0$  & $M_{max}$ & $R_{M_{max}}$  & $R_{M=1.4 M_{\odot}}$  \\
        & (MeV)&(fm$^{-3}$)&              &         &          &(MeV)&  (MeV) & (MeV)  & (M$_{\odot}$) &  (km)  & (km)   \\
\hline
QMC              & -16.4  & 0.15  & 0.00 & 8.6510  & 0.77 & 34.50  & 90.00 & 295 & 2.14  & 11.51    & 13.55  \\
QMC$\omega\rho$ & -16.4  & 0.15  & 0.03 & 9.0078  & 0.77 & 30.92  & 69.17 & 295 & 2.07  & 10.96    & 12.83  \\
\hline
\hline
\end{tabular}
\label{tab:bulk}
\end{table*}

The energy of a static bag describing nucleon $N$ consisting of three quarks in ground state
is expressed as
\begin{equation}
E^{\rm bag}_N=\sum_q n_q \, \frac{\Omega_{q_{N}}}{R_N}-\frac{Z_N}{R_N}
+\frac{4}{3}\,  \pi \, R_N^3\,  B_N\ ,
\label{ebag}
\end{equation}
where $Z_N$ is a parameter which accounts for zero-point motion
of nucleon $N$ and $B_N$ is the bag constant.
The set of parameters used in the present work is determined by enforcing 
stability of the nucleon (here, the ``bag''), much like in \cite{alex09}, 
so there is a single value for 
proton and neutron masses. The effective mass of a nucleon bag at rest
is taken to be $M_N^*=E_N^{\rm bag}.$

The equilibrium condition for the bag is obtained by
minimizing the effective mass, $M_N^*$ with respect to the bag radius
\begin{equation}
\frac{d\, M_N^*}{d\, R_N^*} = 0,\,\,\;\;\; N=p,n.
\label{balance}
\end{equation}
By fixing the bag radius $R_N=0.6$ fm and the bare nucleon mass
$M=939$ MeV the unknowns $Z_N=4.0050668$ and 
$B^{1/4}_N=210.85$MeV are then obtained.
Furthermore, the desired values for the binding energy and
saturation density (Table \ref{tab:bulk}), are achieved by setting $g_\sigma^q=5.9810$, $g_{\omega}=8.9817$, 
where $g_\omega =3g^q_\omega$ and $g_\rho =g^q_\rho$. The meson masses are $m_{\sigma}=550$ MeV,
$m_{\omega}=783$ MeV and $m_{\rho}=770$ MeV. With this
parameterization, some of the bulk properties at saturation density, namely, 
the compressibility, the symmetry energy and the slope of the symmetry
energy, are calculated and the values can be seen in Table \ref{tab:bulk}.  
Other parameter sets are possible, as discussed in \cite{Gramspasta,PandaKrein,Santos-09,panda}.
The values given in the first line of Table \ref{tab:bulk} are very close to the most accepted
values (see \cite{dutra14,oertel16}, for instance) and $J$ and $L_0$ can be
easily controlled by the inclusion of a $\omega-\rho$ interaction, as discussed
in \cite{EPJA_2014,Prafulla_2012,Cavagnoli_2011}. The larger the value of this
interaction, the lower the values of the symmetry energy and its
slope.  In the present work, we follow the calculations mentioned in \cite{Gramspasta}, where
an $\omega-\rho$ interaction strength 
that results in a symmetry energy equal to 22 MeV at 0.1 fm$^{-3}$ is included,
with a consequent change in the $g_\rho$ coupling constant. The new
values of the symmetry energy and its slope at saturation are also
given in Table \ref{tab:bulk}. 

In a relativistic mean field (RMF) approximation, the total energy density of the nuclear matter reads 
\begin{eqnarray}
\varepsilon = \frac{1}{2}m^{2}_{\sigma}\sigma+
\frac{1}{2}m^{2}_{\omega}\omega^{2}_{0}+
\frac{1}{2}m^{2}_{\rho}\rho^{2}_{03}
+3\Lambda_v g_{\omega}^{2}g_{\rho}^{2}\omega^{2}_{0}\rho^{2}_{03} \nonumber \\
+\sum_{N} \frac{1}{\pi^{2}}\int^{k_{N}}_{0}k^{2}dk[k^{2}+M^{*2}_{N}]^{1/2},
\label{energdens}
\end{eqnarray}

and the pressure is,
%\bigskip
%
\begin{eqnarray}
p = -\frac{1}{2}m^{2}_{\sigma}\sigma+
\frac{1}{2}m^{2}_{\omega}\omega^{2}_{0}+
\frac{1}{2}m^{2}_{\rho}\rho^{2}_{03}
+\Lambda_v g_{\omega}^{2}g_{\rho}^{2}\omega^{2}_{0}\rho^{2}_{03} \nonumber \\
+\sum_{N} \frac{1}{\pi^{2}}\int^{k_{N}}_{0}k^{4}dk/[k^{2}+M^{*2}_{N}]^{1/2}.
\label{press}
\end{eqnarray}

The vector mean field $ \omega_0 $ and $ \rho_{03} $ are determined
through
\begin{equation}
\omega_0 =\frac{g_\omega (n_p +n_n)}{m^{*^{2}}_{\omega}}, \;
\rho_{03}=\frac{g_\rho (n_p -n_n)}{2 m^{*^{2}}_{\rho}},
\label{vector}
\end{equation}
where
\begin{equation}
n_B=n_p+n_n = \sum_N \frac{2 k_{N}^3}{3 \pi ^2}, \quad N=p,n.
\end{equation}
is the baryon density, and the effective masses of the meson fields
are 
\begin{equation}
m^{{*}^{2}}_{\omega}=m^{2}_{\omega}+2\Lambda_v
g_{\omega}^{2}g_{\rho}^{2}\rho^2_{03}
\end{equation}
and
\begin{equation}
m^{{*}^{2}}_{\rho}=m^{2}_{\rho}+2\Lambda_v g_{\omega}^{2}g_{\rho}^{2}\omega^2_{0}. 
\end{equation}

Finally, the mean field $\sigma$ is fixed by imposing that
\begin{equation}
\frac{\partial \varepsilon}{\partial \sigma}=0.
\label{condition}
\end{equation}
Our interest lies on stellar matter 
in  $\beta$-equilibrium conditions, i.e.,
\begin{equation}
\mu_p=\mu_n-\mu_e, \quad \mu_e=\mu_\mu.
\end{equation}
Charge neutrality requires that
\begin{equation}
n_p=n_e + n_\mu,
\end{equation}
and these two conditions imply that a free gas of leptons (electrons and muons) have to be added to the energy density and pressure of the system. 

\vspace{0.2cm}

\begin{itemize}
    \item {\bf Pasta phases}
\end{itemize}

\vspace{0.2cm}

We construct the pasta phases within the QMC and QMC$\omega\rho$ models
using the coexisting phases method \cite{Maruyama-05,Avancini-08}. 
For a given total density $n_B$ the pasta 
structures are built with different geometrical forms, usually called   
sphere (bubble),  cylinder (tube), and slab, in three, two,
and one dimensions, respectively.
This is achieved from the Gibbs conditions,
that impose that both phases have the same pressure, proton
and neutron chemical potentials. For stellar matter, the following equations
must be solved simultaneously
\begin{equation}
P^I=P^{II},
\label{p1e2}
\end{equation}
\begin{equation}
\mu_p^I=\mu_p^{II},
\end{equation}
\begin{equation}
\mu_n^I=\mu_n^{II},
\end{equation}
\begin{equation}
f(n^{I}_p-n^{I}_e)+(1-f)(n^{II}_p-n^{II}_e)=0.
\end{equation}
where $I$ ($II$) represents the high (low) density region, $n_p$ 
is the global proton density and 
$f$ is the volume fraction of the phase $I$, that reads
\begin{equation}
f=\frac{n_B-n_B^{II}}{n_B^I-n_B^{II}}.
\end{equation} 

The hadronic matter energy reads:
\begin{eqnarray}
\varepsilon_{matter}=f\varepsilon^I+(1-f)\varepsilon^{II}+\varepsilon_e.
\label{energymat}
\end{eqnarray}

Adding the surface and Coulomb terms to Eq. (\ref{energymat}) results in the 
total energy density:
\begin{eqnarray}
\varepsilon=\varepsilon_{matter}
  +\varepsilon_{surf}+\varepsilon_{Coul}.
\label{totalener}
\end{eqnarray}

Minimizing $\varepsilon_{surf}+\varepsilon_{Coul}$ with respect to the size of the
droplet/bubble, cylinder/tube or slabs, we obtain \cite{maruyama} 
$\varepsilon_{surf}=2 \varepsilon_{Coul}$ where
\begin{equation}
\varepsilon_{Coul}=\frac{2\alpha}{4^{2/3}}(e^{2}\pi \Phi)^{1/3}\left[ \sigma D(n^{I}_p -n^{II}_p)\right] ^{2/3},
\end{equation}
with $\alpha=f$ for droplets, tubes and slaps, and $\alpha=1-f$ for 
tubes and bubbles.
$\Phi$ is given by
\begin{eqnarray}
\Phi=\left\lbrace\begin{array}{c}
\left(\frac{2-D \alpha^{1-2/D}}{D-2}+\alpha\right)\frac{1}{D+2}, \quad
                  D=1,3 \\
\frac{\alpha-1-\ln \alpha}{D+2},  \quad D=2 \quad 
\end{array} \right . 
\end{eqnarray}
$\sigma$ is the surface tension, which measures the energy per area necessary
to create a planar interface between the two regions and is calculated using
an adapted geometric approach \cite{Gramspasta}.

Notice that the pasta phase is only present at the low density regions of the neutron
stars and in this region muons are not present, although they are present in the EOS that describes the homogeneous region.  

\subsection{MQMC model}

The Lagrangian density of the modified quark-meson coupling (MQMC) model, extended in comparison 
with its standard form~\cite{jpg,npa} in order to take into account asymmetric nuclear 
matter~\cite{rnm,hss,prc16}, is given by
\begin{eqnarray}
\mathcal{L}_{\mbox{\tiny MQMC}} &=& \overline{\psi}_q[i\gamma^\mu\partial_\mu - m_q - 
U(r)]\psi_q 
+ g_\sigma^q\sigma\overline{\psi}_q\psi_q \nonumber
\\
&-& g_\omega^q\overline{\psi}_q\gamma^\mu\omega_\mu\psi_q 
- \frac{g_\rho^q}{2}\overline{\psi}_q\gamma^\mu\vec{\rho}_\mu\vec{\tau}\psi_q \nonumber
\\
&+& \frac{1}{2}(\partial^\mu \sigma \partial_\mu \sigma 
- m^2_\sigma\sigma^2)
-\frac{1}{4}F^{\mu\nu}F_{\mu\nu}
+ \frac{1}{2}m^2_\omega\omega_\mu\omega^\mu \nonumber
\\
&-&\frac{1}{4}\vec{B}^{\mu\nu}\vec{B}_{\mu\nu}
+ \frac{1}{2}m^2_\rho\vec{\rho}_\mu\vec{\rho}^\mu,
\label{dl}
\end{eqnarray}
The antisymmetric field tensors $F_{\mu\nu}$ and $\vec{B}_{\mu\nu}$ are given by 
$F_{\mu\nu}=\partial_\nu\omega_\mu-\partial_\mu\omega_\nu$ and 
$\vec{B}_{\mu\nu}=\partial_\nu\vec{\rho}_\mu-\partial_\mu\vec{\rho}_\nu$.

In the original QMC model~\cite{guichon} introduced previously, quarks interact each other through an MIT-like 
bag potential and the nucleon is described as a bag composed of three quarks. Each 
set of three quarks interact with another one through meson exchanges, as already mentioned. The same kind of 
nucleon interaction is also present in the MQMC model. However, the interaction between 
quarks inside the nucleon is taken into account via a confining harmonic oscillator 
potential given by $U(r) = \frac{1}{2}(1+\gamma^0)V(r)$, with $V(r)=ar^2+V_0$, instead of the 
bag-like treatment. The potential intensity and depth are related, respectively, to the 
$a$ and $V_0$ constants. 

From the Euler-Lagrange equations applied to the Lagrangian density in Eq.~(\ref{dl}), 
one can write the Dirac equation for the quarks as
\begin{align}
\left\lbrace\vec{\alpha}\cdot\vec{k}+\gamma^0\left[m_q-V_\sigma+\frac{V(r)}{2}\right] + 
\frac{V(r)}{2} + V_\omega + \frac{\tau_zV_\rho}{2}\right\rbrace\psi_q  \nonumber\\
=\varepsilon_q\psi_q
\end{align}
with
\begin{eqnarray}
\vec{\alpha}\cdot \vec{k}=\left(
\begin{array}{cc}
               0          &  \vec{\sigma}\cdot \vec{k}\\
\vec{\sigma}\cdot \vec{k} &  0
\end{array}
\right), \quad \psi_q = \left(
\begin{array}{c}
\varphi \\
\chi
\end{array}
\right),
\end{eqnarray}
and
\begin{eqnarray}
V_\sigma = g_\sigma^q\sigma,\quad V_\omega = g_\omega^q\omega_0,\quad V_\rho = 
g_\rho^q\rho_{03},
\end{eqnarray}
where $\sigma$, $\omega_0$ and $\rho_{03}$ are the classical meson fields in the mean-field 
approximation. 

If we write the small component ($\chi$) of the Dirac field in terms of the larger one 
($\varphi$), and replace it back into one of the coupled equations for $\chi$ and 
$\varphi$, we will find
\begin{align}
\left[\frac{k^2}{2(\varepsilon_q^* + 
m_q^*)}+\frac{a}{2}r^2\right]\varphi(\vec{r}) = 
\frac{\varepsilon_q^* - m_q^* - V_0}{2}\varphi(\vec{r}),
\label{schrodinger}
\end{align}
with
\begin{eqnarray}
\varepsilon_q^* = \varepsilon_q - V_\omega - \frac{1}{2}\tau_zV_\rho
\end{eqnarray}
and
\begin{eqnarray}
m_q^*=m_q - V_\sigma.
\end{eqnarray}

The Schr\"odinger equation of the tridimensional harmonic oscillator is recognized in 
Eq.~(\ref{schrodinger}), with the lowest order energy identified in the right-hand side. 
Such an identification gives rise to 
\begin{eqnarray}
\varepsilon_q^* - m^*_q - V_0 = 3\sqrt{\frac{a}{\varepsilon_q^*+m^*_q}},
\label{cubic}
\end{eqnarray}
by taking into account that $\frac{3}{2}\omega=\frac{\varepsilon_q^* - m^*_q - 
V_0}{2}$ with $\omega=\sqrt{\frac{a}{\varepsilon_q^*+m^*_q}}$, in 
units of $\hbar = c = 1$.

Since the center of mass motion of the composite nucleon in the MQMC model is also bound by the 
harmonic potential between the quarks, corrections in the nucleon wave function must to be taken 
into account if we consider the composite state as a translationally invariant one. Here, we 
follow the procedure used in Ref.~\cite{npa} in order to extract center of mass effects from 
nucleon observables. Firstly, we obtain the center of mass energy, given by~\cite{npa}
\begin{eqnarray}
\varepsilon_{\rm cm} = \frac{3}{2}\frac{\alpha}{(\varepsilon_q^* + m_q^*)} 
\frac{(3+23\beta/6)}{(1+3\beta/2)^2},
\label{ecm}
\end{eqnarray}
with 
\begin{eqnarray}
\alpha = \sqrt{a}(\varepsilon_q^* + m_q^*)^{1/2}
\end{eqnarray}
and
\begin{eqnarray}
\beta = \frac{\alpha}{(\varepsilon_q^* + m_q^*)^2} = \sqrt{a}(\varepsilon_q^* + m_q^*)^{-3/2}.
\end{eqnarray}

The effective nucleon mass in the medium as the center of mass corrected energy of the three 
independent quarks is taken into account is then expressed as~\cite{npa}
\begin{eqnarray}
M_N^* = 3\varepsilon_q^* - \varepsilon_{\rm cm}, 
\end{eqnarray}
and the mean squared nucleon radius, also corrected for center of mass effects, is written 
as~\cite{npa}
\begin{eqnarray}
\left<r_N^2\right> = \frac{1+5\beta/2}{\alpha(1+3\beta/2)}.
\end{eqnarray}
The harmonic oscillator parameters $a$ and $V_0$ are determined by imposing the vacuum values for 
$M_N^*$ and $\left<r_N^2\right>$. Here we adopt $M_N^*(n=0)=939$~MeV and 
$\left<r_N^2\right>(n=0)=0.8^2$~fm$^2$.

The equations of state (EoS) and field equations of the MQMC model are given as in the QMC one, by taking 
$\Lambda_v=0$. More specifically, the energy density and pressure of the MQMC model are given by Eqs.~(\ref{energdens}) 
and (\ref{press}), respectively. The mean fields $\omega_0$, $\rho_{03}$ and $\sigma$ are obtained as indicated in 
Eqs.~(\ref{vector}) and (\ref{condition}), all of them with the restriction that the $\Lambda_v$ parameter is set equal to zero.

The free parameters $G_\sigma^{q2}\equiv (g_\sigma^q/m_\sigma)^2$ and $G_\omega^2\equiv (g_\omega/m_\omega)^2$ are 
found by imposing the nuclear matter saturation at $n=n_0=0.15$~fm$^{-3}$ with a binding energy of $B/A=16$~MeV. 
Finally, $G_\rho^2\equiv (g_\rho/m_\rho)^2$ is determined by fixing a particular value for $J\equiv\mathcal{S}(n=n_0)$, 
with the symmetry energy given by~\cite{hss,prc16}
\begin{eqnarray}
\mathcal{S} &=& \frac{k_F^2}{6(k_F^2 + M_N^{*2})^{1/2}} + \frac{1}{8}G_\rho^2n,
\end{eqnarray}
with $k_F$ being the Fermi momentum. The input free parameter $m_q$ is used to control the incompressibility at the 
saturation density $K_0=K(n=n_0)$, with $K=9\partial p/\partial n$. Here we restrict the MQMC model to present the same $J$ and $K_0$ values as those from the QMC models in Sec.~\ref{qmc}, see Table~\ref{tab:bulk}. Such parametrizations are named as  MQMC1 and MQMC2. We also generate a third one, namely, MQMC3 in which one has $J=25$~MeV. Notice that all these parametrizations present $J$ and $K_0$ values inside the ranges of $25\,\mbox{MeV}\leqslant J \leqslant 35\,\mbox{MeV}$~\cite{dutra14} and $250\,\mbox{MeV}\leqslant K_0 \leqslant 315\,\mbox{MeV}$~\cite{stone}, respectively.
\begin{table*}
\centering
\caption{Nuclear matter and stellar properties obtained from the parametrizations of the MQMC model. The free parameters are also given. $G_\sigma^{q2}$, $G_\omega^2$ and $G_\rho^2$ are given in $10^{-5}$MeV$^{-2}$. For all parametrizations, one has $B/A=-16.4$~MeV and $n_0=0.15$~fm$^{-3}$.}
\begin{tabular}{lccccccccccccc}
\hline\hline
Model & $m_q$ & $K_0$ & $M^*_N/M_N$ & $J$ & $L_0$ & $a$ & $V_0$ & $G_\sigma^{q2}$ & $G_\omega^2$ & $G_\rho^2$ & $M_{max}$ & $R_{M_{max}}$  & $R_{M=1.4 M_{\odot}}$ \\
& (MeV) & (MeV) & - & (MeV) & (MeV) & (fm$^{-3}$) & (MeV) &  &  & & ($M_{\odot}$) & (km) & (km) \\
\hline
MQMC1 & $210.61$ & $295$  & $0.84$  & $34.50$ & $93.20$ & $0.95$ & $-92.27$ & $5.13$ & $8.33$ & $14.67$ & $1.97$ & $11.43$ & $13.55$\\
MQMC2 & $210.61$ & $295$  & $0.84$  & $30.92$ & $82.46$ & $0.95$ & $-92.27$ & $5.13$ & $8.33$ & $12.18$ & $1.97$ & $11.34$ & $13.32$\\
MQMC3 & $210.61$ & $295$  & $0.84$  & $25.00$ & $64.70$ & $0.95$ & $-92.27$ & $5.13$ & $8.33$ & $ 8.08$ & $1.97$ & $11.18$ & $12.94$\\
\hline\hline
\end{tabular}
\label{tabmqmc}
\end{table*}

For the stellar matter calculations, we proceed as described in Sec.~\ref{qmc} concerning the $\beta$-equilibrium 
conditions on the chemical potentials and densities. In particular, the nucleon chemical potentials in the MQMC model 
are given by,
\begin{align}
\mu_{p,n} = (k_F^2 + M_N^{*2})^{1/2} + G_\omega^2(n_p+n_n)\,\pm \frac{1}{4}G_\rho^2(n_p-n_n)
\end{align}
with the upper (lower) sign for protons (neutrons). The nuclear matter and stellar properties, along with the free parameters obtained from the parametrizations of the MQMC model, are given in Table.~\ref{tabmqmc}.

\section{GW170817 constraints}
\label{GW}

The gravitational Love number depends directly on the detailed structure of the neutron star (NS). Therefore the observation of Love numbers can offer us paramount information on the NS composition. In fact this physics has triggered intense research recently \cite{PhysRevC.87.015806,PhysRevC.98.035804,PhysRevC.98.065804,PhysRevC.95.015801}.

When one of the neutron stars in a binary system
gets close to its companion just before merging, a mass
quadrupole develops as a response to the tidal field induced by the companion. This is known as tidal polarizability \cite{PhysRevD.80.084035,PhysRevD.80.084018} and can be used to constrain neutron star macroscopic properties \cite{PhysRevD.77.021502}, which in turn, are obtained from appropriate equations of state (EOS). 

In a binary system the induced quadrupole moment $Q_{ij}$ in one neutron star due to the external tidal field ${\cal E}_{ij}$ created by a companion compact object can be written as~\cite{PhysRevD.77.021502},
\begin {equation}
Q_{ij} = -\lambda {\cal E}_{ij},
\end{equation}
where, $\lambda$ is the tidal deformability parameter, which can be expressed in terms of 
dimensionless $l = 2$ quadrupole tidal Love number $k_2$ as
\begin{equation}
\label{tidal}
\lambda= \frac{2}{3} {k_2}R^{5}.
\end{equation}

 To obtain $k_{2}$, we have to simultaneously solve the TOV equations and find the value of $y$ from the following differential equation
\begin{equation}
r \frac{dy}{dr} + y^2 + y F(r) + r^2Q(r)=0,
\label{ydef}
\end{equation}
with its coefficients given by
\begin{equation}
F(r) = \frac{r - 4\pi r^3(\varepsilon - p)}{r-2m}
\end{equation}
and
\begin{align}
Q(r)=&\frac{4\pi r \left(5\varepsilon + 9p + \frac{(\varepsilon + p)}{\partial p/\partial \varepsilon}-\frac{6}{4\pi r^2}\right)}{r-2m} 
\nonumber\\ 
& - 4\left( \frac{m+4\pi r^3 p}{r^2-2mr} \right)^2, 
\end{align}
where $\varepsilon$ and $p$ are the energy density and pressure profiles inside the star. Then we can compute the Love number $k_2$, which is given by 
\begin{align}
k_2 =&\frac{8C^5}{5}(1-2C)^2[2+2C(y_R-1)-y_R]\times
\nonumber\\
&\Big\{2C [6-3y_R+3C(5y_R-8)]
\nonumber\\
&+4C^3[13-11y_R+C(3y_R-2) + 2C^2(1+y_R)]
\nonumber\\
&+3(1-2C)^2[2-y_R+2C(y_R-1)]{\rm ln}(1-2C)\Big\}^{-1},
\label{k2}
\end{align}
where $y_{R}=y(r=R)$, $C=M/R$ is the compactness of the star and $R$ is its radius. 

The tidal deformability $\Lambda$ (i.e., the dimensionless version of 
$\lambda$) is connected with the compactness parameter $C$ through 
\begin{equation}
\Lambda= \frac{2k_2}{3C^5}. 
\label{dtidal}
\end{equation}  

{\bf In the next section, $\Lambda_1$ and $\Lambda_2$ refer to the the values of each one of the neutron star in the binary system.}

\subsection{The inner and outer crust effects on the tidal polarizability}

\begin{figure}[!htb]
\includegraphics[scale=0.34]{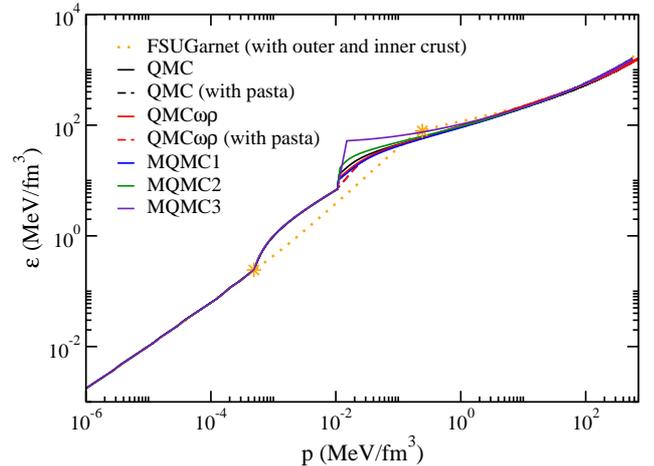}
\caption{Stellar matter EOSs described by QMC and MQMC models in comparison with the FSUGarnet model\cite{jorge2018}. The region in between the two stars in the orange curve represents the inner crust. Before (after) that region, one has the outer crust (liquid core).}
\label{totalEOS}
\end{figure}

It is obviously expected that the crust thickness and constitution affect the second Love number and consequently, the tidal polarizability. The neutron star crust is usually divided into two different parts: the outer crust and the inner crust. In \cite{jorge2018}, the authors investigated the impact of the crust by considering simple expressions. For the outer crust, the region where all neutrons are bound to finite nuclei, a crystal lattice calculation that depends on the masses of different nuclei was performed. For the inner crust, where the pasta phase is expected to exist, the authors used a polytropic EOS that interpolates between the homogeneous core and the outer crust. The conclusion of their work is that, for a fixed compactness, the second Love number is sensitive to the inner crust, but as the tidal polarizability scales as the fifth power of the compactness parameter, the overall impact is minor. In the present work, we use the BPS EOS \cite{bps1971} for the outer crust and the pasta phase for the inner crust and test their effects on the deformability of the NS. We can see the differences between the total equations of state (outer crust + inner crust + liquid core) presented here and the one used in \cite{jorge2018} in Fig.\ref{totalEOS}
One can see that the outer crusts are coincident and the liquid core of all EOS are very similar. Hence, most of the differences reside on the inner crust and around the crust-core transition region. Notice that the pasta phase is present only in the QMC and QMC$\omega\rho$ models.
 
In the next section, we obtain the results for the second Love number and the tidal deformability and discuss them. 

\section{Neutron star cooling}
\label{cooling}

We now turn our attention to the thermal evolution of neutron stars whose composition is described by the models discussed in this manuscript.
The cooling of neutron stars is governed by the emission of neutrinos from their core, and of photons from the surface. All thermal properties of the neutron star from the neutrino emission to the heat transport, depend on their microscopic composition, as different compositions lead to different processes, or alter the rate at which certain processes take place. Evidently the specific heat and thermal conductivity also depend on the microscopic composition of the star. Furthermore global and macroscopic properties of the star, such as size, spin, crystalline structure (of the crust) also play a major role in the thermal evolution of the star. Cooling studies are thus a superb way of bridging the gap between the micro and macroscopic realms of neutron stars. 

In recent years there has been great advances in the observation of thermal properties of compact objects \citep{Beloin2016a,SafiHarb2008,Zavlin1999,Pavlov2002,Mereghetti1996,Zavlin2007,Pavlov2001,Gotthelf2002,McGowan2004,Klochkov2015,McGowan2003,McGowan2006,Possenti1996,Halpern1997,Pons2002,Burwitz2003,Kaplan2003,Zavlin,Ho2015}.
%Such developments have ignited a great interest in the research of the thermal properties of neutron stars as well as their thermal evolution, as these are intrinsically connected to the microscopic and macroscopic physics of the star. This fact allows us to use thermal evolution simulation, together with observations, as a way of probing the interiors of such stars. 

%The cooling of neutron stars is governed by the emission of neutrinos from the star interior as well as photon emissions from its surface. 
The equations that describe thermal energy balance and transport inside a spherically symmetric general relativistic star are given in
\cite{2006NuPhA.777..497P,1999Weber..book,1996NuPhA.605..531S}, with ($G = c = 1$) 
\begin{eqnarray}
  \frac{ \partial (l e^{2\phi})}{\partial m}& = 
  &-\frac{1}{\rho \sqrt{1 - 2m/r}} \left( \epsilon_\nu 
    e^{2\phi} + c_v \frac{\partial (T e^\phi) }{\partial t} \right) \, , 
  \label{coeq1}  \\
  \frac{\partial (T e^\phi)}{\partial m} &=& - 
  \frac{(l e^{\phi})}{16 \pi^2 r^4 \kappa \rho \sqrt{1 - 2m/r}} 
  \label{coeq2} 
  \, .
\end{eqnarray}
In eqs. (\ref{coeq1})-(\ref{coeq2}) $l$ is the luminosity, $\phi$ is the metric function, $m$ is the mass as a function of the radial distance $r$ and $T$ is the temperature. The equations above also depend on microscopic properties such as the neutrino emissivity $\epsilon_\nu$, the specific heat $c_v$ and the thermal conductivity $\kappa$. A comprehensive review of the details of such quantities can be found in references \citep{Yakovlev2000}.

One also needs the appropriate boundary conditions for a complete solution of eqs. (\ref{coeq1})-(\ref{coeq2}) and these are given by the heat flow at the center of the star, which obviously needs to vanish, as well as surface conditions which connect the stellar surface luminosity to the heat flowing from the mantle to the star atmosphere. The atmosphere conditions may depend on several circumstances such as, surface magnetic field and/or heavier elements due to accreted matter at the neutron star formation \citep{Gudmundsson1982,Gudmundsson1983,Page2006}. In this work we consider the traditional scenario, as described in the aforementioned references, and only briefly consider the possibility of accreted matter in the atmosphere. 

In the next section, we numerically solve eqs.~(\ref{coeq1})-(\ref{coeq2}) for a set of stars of each model, spanning low to high mass neutron stars, and then confront our results to observed data and draw our conclusions.

\section{Results and Discussion}
\label{results}

In the present section we study the effects of the different versions of the QMC and MQMC models presented earlier on quantities related to gravitational wave observables and
on the neutron star cooling.

\subsection{GW170817 constraint results}

We next show the results of the tidal deformability $\Lambda$ and the Love number $k_2$ obtained from the QMC/QMC$\omega\rho$ models, with and without pasta phases, and the three versions of the MQMC.

\begin{figure}[!htb]
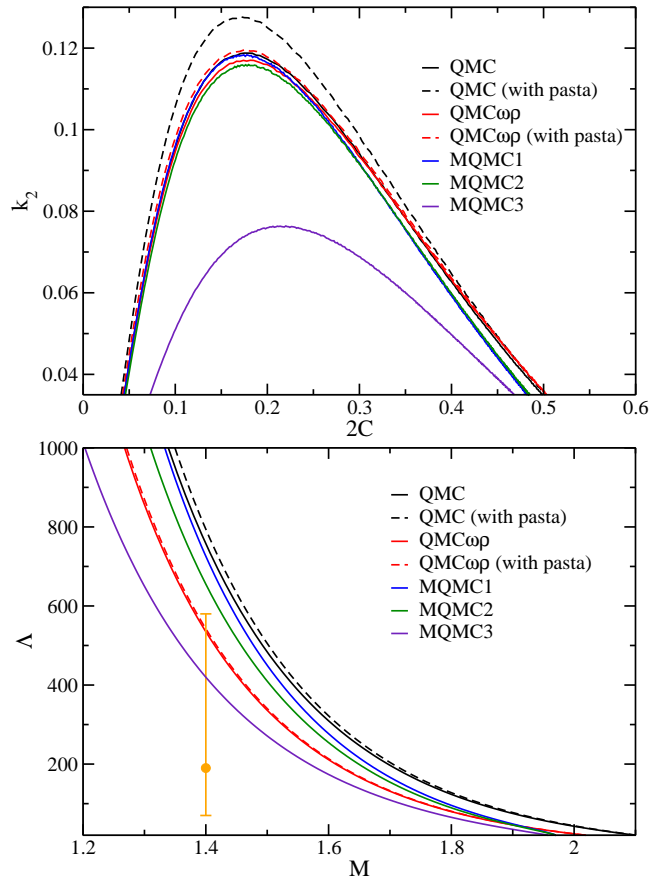

\centering
\includegraphics[scale=0.34]{k2-2.eps}
\includegraphics[scale=0.34]{lm.eps}
\vspace{-0.2cm}
\caption{(top) $k_2$ as a function of the neutron star compactness, and (bottom) $\Lambda$ as a function of $M$. Full circle: recent result of $\Lambda_{1.4}=190_{-120}^{+390}$ obtained by LIGO and Virgo Collaboration~\cite{PRL121_161101}.} 
\label{k2lambdaQMC}
\end{figure}

In Fig.~\ref{k2lambdaQMC}(top) we display the Love number $k_2$ as a function of the compactness of the neutron star. In Fig.~\ref{k2lambdaQMC}(bottom) we can see the variation of $\Lambda$ with the mass of the star. As already pointed out in \cite{jorge2018}, the curves do not collapse into one single curve because of their different dependence on $y$. When the same models are used to compute the tidal polarizability, their differences are enhanced and only two models, namely, QMC$\omega\rho$ and MQMC3 give results that are consistent with LIGO results for the canonical star. The pasta phase plays a very modest role and its influence is practically unnoticed, what can be seen if one compares the curves obtained from QMC with and without pasta and QMC$\omega\rho$ with and without pasta.

\begin{figure}[!htb]
\includegraphics[scale=0.34]{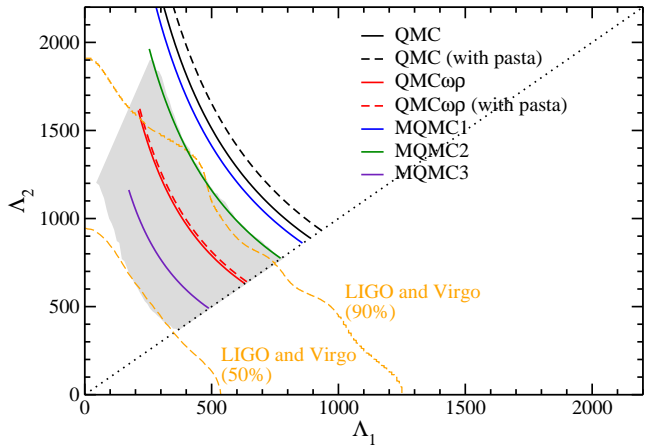}
\caption{Tidal deformabilities obtained from the QMC, QMC$\omega\rho$ and MQMC models for both components of the binary system related to the GW170817 event. The confidence lines (50$\%$ and 90$\%$) are the recent results of LIGO and Virgo collaboration taken from Ref. \cite{PRL121_161101}. The dashed region represents
the results obtained with consistent relativistic mean field models in \cite{OdilonGW170817}.}
\label{lambda}
\end{figure}

In Fig. \ref{lambda} we show the tidal deformabilities of each neutron star in the binary system. $\Lambda_1$ is associated to the neutron star with mass $m_1$, which corresponds to the integration of every EoS in the range $1.37 \leqslant m/M_{\odot}  \leqslant1.60$ obtained from GW170817. The mass $m_2$ of the companion star is determined by solving $ \mathcal{M}_c = 1.188 M_{\odot} = \frac{(m_1 m_2)^{3/5}}{(m_1+m_2)^{1/5}}$ \cite{PRL.119.161101}. We notice that only the QMC$\omega\rho$ model (with and without pasta phases), and the MQMC3 parametrization show values in between the confidence lines. 

\subsection{Cooling results}

\begin{figure}[!htb]
\includegraphics[scale=0.34]{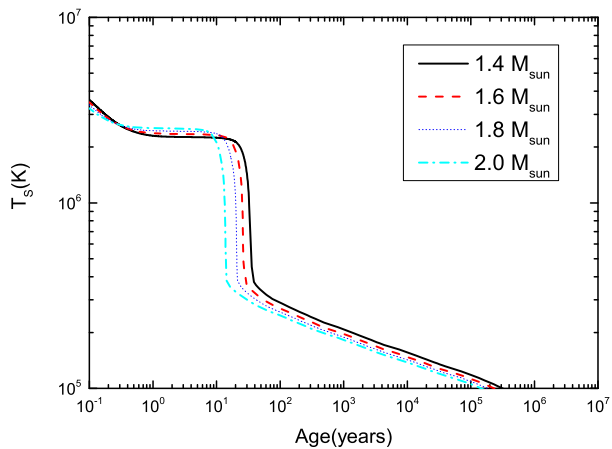}
\caption{Thermal evolution of neutron stars described by the QMC model. Y-axis represent the redshifted surface temperature and x-axis the age. Each curve represents the thermal evolution of a neutron star with a different mass. All objects exhibit fast cooling due to prominent DU process in their core.}
\label{coolQMC1}
\end{figure}

We now present the results of our thermal evolution studies, obtained by the numerical solution of eqs.~(\ref{coeq1}) and (\ref{coeq2}). We begin by showing the cooling of a set of stars covering a wide range of masses for the QMC model. Those are shown in Fig.~\ref{coolQMC1}.

As shown in Fig.~\ref{coolQMC1} all stars exhibit fast cooling, characterized by a sharp drop in their surface temperature at the age of $\sim 100$ years. Such fast cooling is a manifestation of a prominent presence of the direct Urca process (DU) in the stellar core, which is indeed the case for these stars. The DU process is extremely efficient in exhausting the star thermal energy, leading to the manifested fast cooling \cite{Prakash1992}. This, in turn indicates that such thermal evolution is mostly incompatible with observed data (unless the DU process is suppressed), as we will discuss 
in the following.

\begin{figure}[!htb]
\includegraphics[scale=0.34]{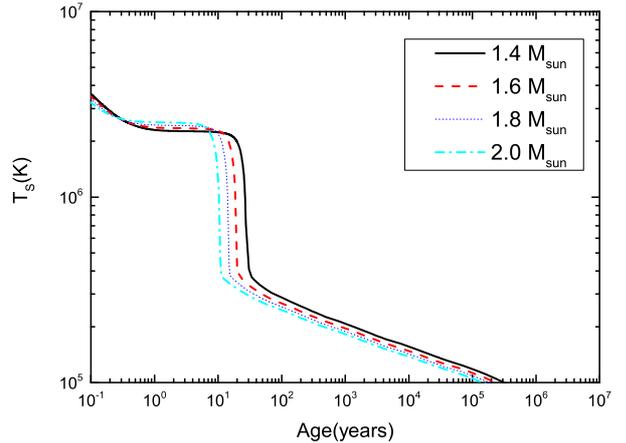}
\caption{Same as in Fig.~\ref{coolQMC1} but for stars with pasta phase in the inner crust.
\label{coolQMC_pasta1}}
\end{figure}

We now show in Fig.~\ref{coolQMC_pasta1}  the thermal evolution of QMC neutron stars with the pasta phase describing the inner crust.

We can see that the pasta phase has little effect on the overall cooling behavior of the star. This is to be expected as the pasta phase occupies only a small region. A more careful analysis lead us to the conclusion that the pasta phase had the minor effect of aiding in the core-crust thermalization, leading to a core-crust thermal relaxation on average $\sim 3.5$ years faster. This seems reasonable, as the pasta phase smooths the core-crust transition, thus, one can expect that will also smooth out heat propagation in between these regions. We note, however, that these are rough estimates and a more detailed calculation is warranted, one in which the thermal and conductivity properties of the pasta phase are explored in more details. 

\begin{figure}[!htb]
\includegraphics[scale=0.34]{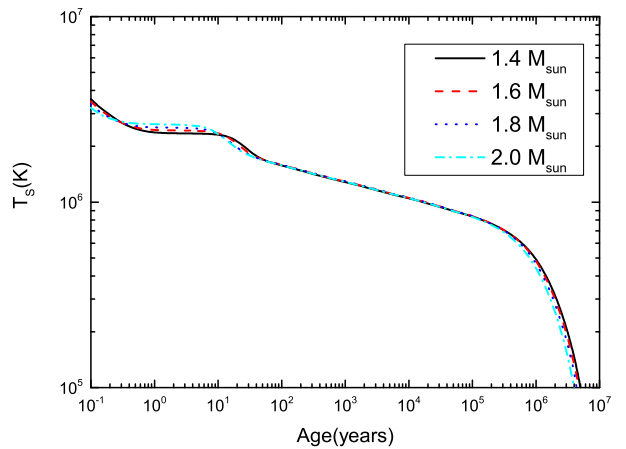}
\caption{Surface temperature as a function of age for stars within the QMC$\omega \rho$ model. Each curve represents the cooling of stars with the indicated mass. \label{cool_QMCwr}}
\end{figure}

\begin{figure}[!htb]
\includegraphics[scale=0.34]{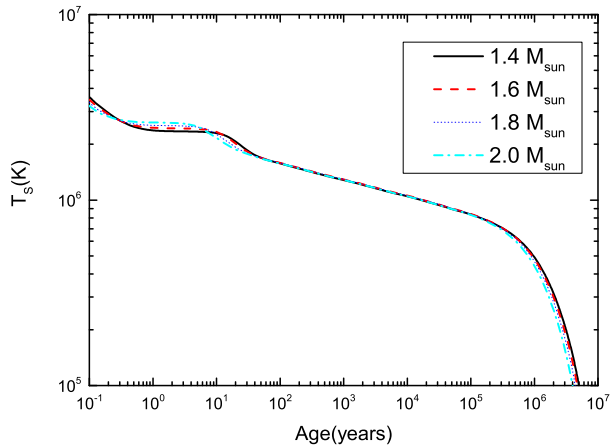}
\caption{Same as Fig.~\ref{cool_QMCwr} but for stars with a pasta phase. \label{cool_QMCwrpasta}}
\end{figure}

We now repeat the calculations above, but for the QMC$\omega \rho$ model, without pasta (Fig.~\ref{cool_QMCwr}) and with pasta (Fig.~\ref{cool_QMCwrpasta}).

We can see that, as happens in the previous case (QMC), the pasta phase has little effect on the overall cooling - serving only to delay the thermalization by a few years. Differently than the previous model, however, in this case all stars exhibit slow cooling, meaning that there is no sharp drop in temperature when the core and crust become thermally coupled. The reason for that is that in this particular model the proton fraction at the core of the star, even at larger densities of heavier stars, is low enough to prevent the DU process from taking place, thus leading to a substantially slower thermal evolution.

\begin{figure}[!htb]
\includegraphics[scale=0.34]{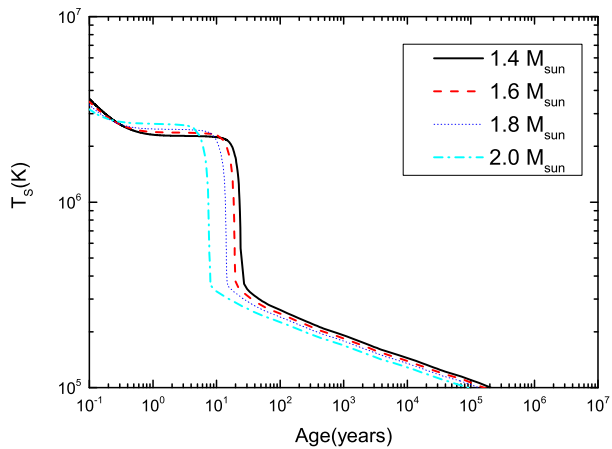}
\caption{Thermal evolution of different neutron stars under the MQMC1 model \label{coolMQMC1}}
\end{figure}

\begin{figure}[!htb]
\includegraphics[scale=0.34]{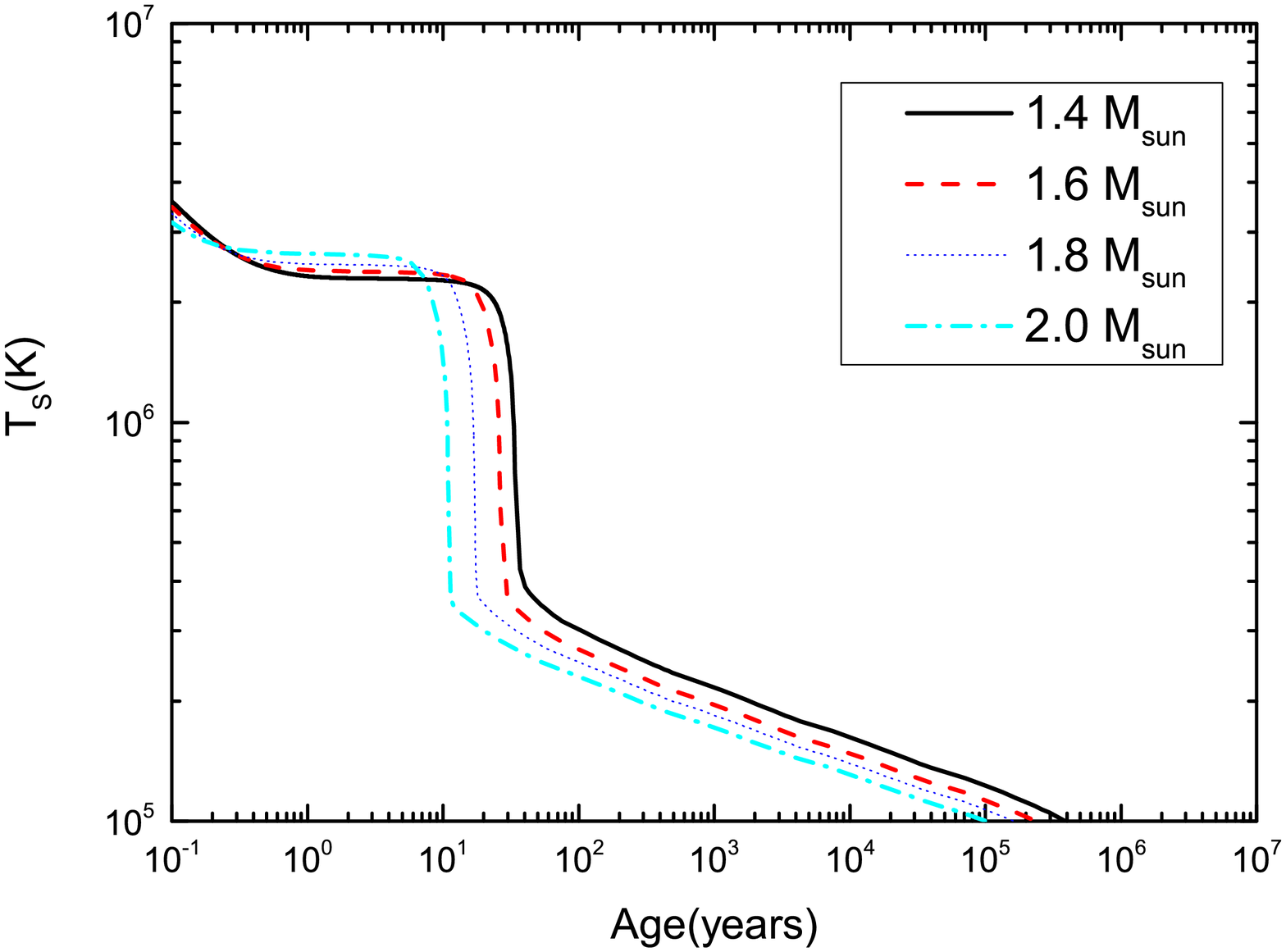}
\caption{Same as Fig.~\ref{coolMQMC1} but for the MQMC2 model \label{coolMQMC2}}
\end{figure}

\begin{figure}[!htb]
\includegraphics[scale=0.34]{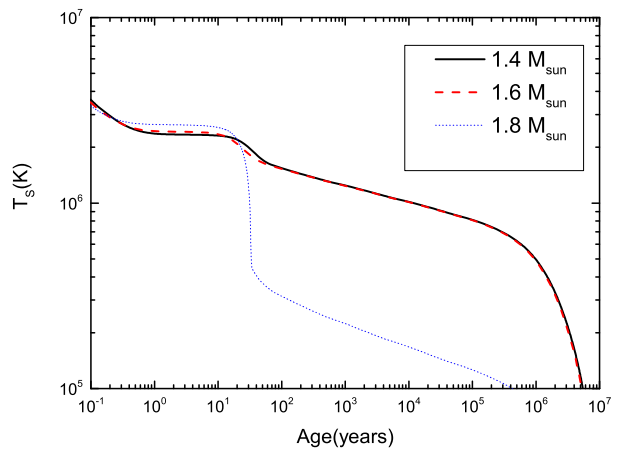}
\caption{Same as Fig.~\ref{coolMQMC1} but for the MQMC3 model \label{coolMQMC3}}
\end{figure}

Now we investigate the cooling of neutron stars described by the MQMC model in order to see if the different prescription used in the EoS calculations will play any role in the cooling. We show in Figs.~\ref{coolMQMC1} - \ref{coolMQMC3} the cooling of neutron stars of different masses calculated for the different MQMC models studied. 

We can see by the results shown in Figs.~\ref{coolMQMC1} - \ref{coolMQMC3} that the overall cooling behavior associated with each model is, qualitatively, the same. For the MQMC models 1 and 2 we see that all stars exhibit fast cooling, with MQMC 3 displaying slow cooling (except for high mass stars). 

It is worth mentioning that we see a clear correlation between the slope of the nuclear symmetry energy ($L_0$) and a fast/slow cooling behavior, with stars with higher values of $L_0$ having faster cooling (QMC, MQMC1 and MQMC2) whereas lower values of $L_0$ leading to slower cooling scenario (QMC$\omega \rho$ and MQMC3).
%These results indicate that the thermal behavior is qualitatively the same for both models studied. 

\begin{figure}[!htb]
\includegraphics[scale=0.34]{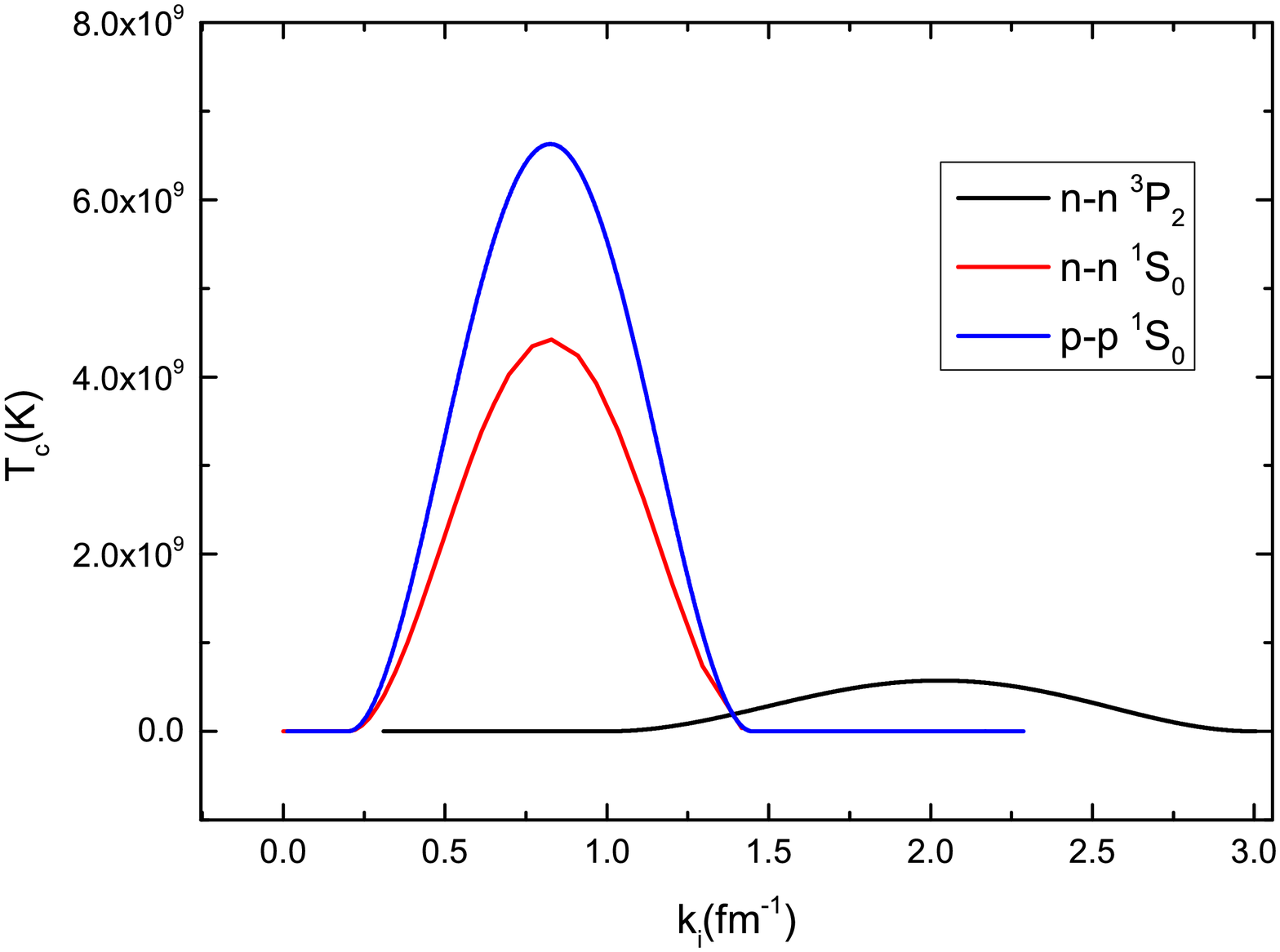}
\caption{Critical temperature for the different pairing taking place inside of the neutron star. $k_i$ stands for the fermi momentum of the neutron and proton, according to the pairing in question. \label{TCall}}
\end{figure}

So far in our thermal investigation we have completely ignore effects of pairing, as we were interested in probing possible differences in the thermal behavior of the different models studied, and the inclusion of pairing could potentially murk the results. As it turns out all models behave similarly, with major differences connected with the different values of the symmetry energy slope rather than more specific minutia about the models. We now need to make sure that our model is in agreement with observed data, or at the very least is comparable to other results in the recent literature \cite{Negreiros2015a,Negreiros2015b,Beloin2018,Negreiros2018a,Raduta2018} - for that we need to include pairing. Pairing in neutron stars may potentially occur in three different manners, neutron-neutron singlets ($^1S_0$), neutron-neutron triplet ($^3P_2$) and proton-proton singlet ($^1S_0$). Neutron-neutron singlets are known to form mainly in the neutron-free region of the crust, whereas its triplet counterpart should form mainly in the lower densities of the core. As for the proton-proton pairs there is still great uncertainty as to how deep into the star it may take place. For a review of pairing in neutron stars we refer the reader to \cite{Yakovlev2000}. In this work we adopt a fairly standard pairing scenario, similar to the CCDK model of reference \cite{Page2004}. Below we show in (Figs.~\ref{TCall}) the critical temperature as a function of neutron and proton Fermi momenta that were used to calculate the pairing effects in the neutron star cooling processes (suppression of the DU process, appearance of Pair-Breaking-Formation (PBF) process near $T_c$, and the modification of the specific heat of paired particles).

As discussed above, the presence of pasta does not change (qualitatively) the cooling behavior of the star, thus henceforth we will consider only the QMC with the pasta phase, as we believe that this is a more appropriate description for the core-crust transition. We now revisit the results shown in Figs.~\ref{coolQMC_pasta1} - \ref{coolMQMC3} taking into account the superfluidity model described above. 

\begin{figure}[!htb]
\includegraphics[scale=0.34]{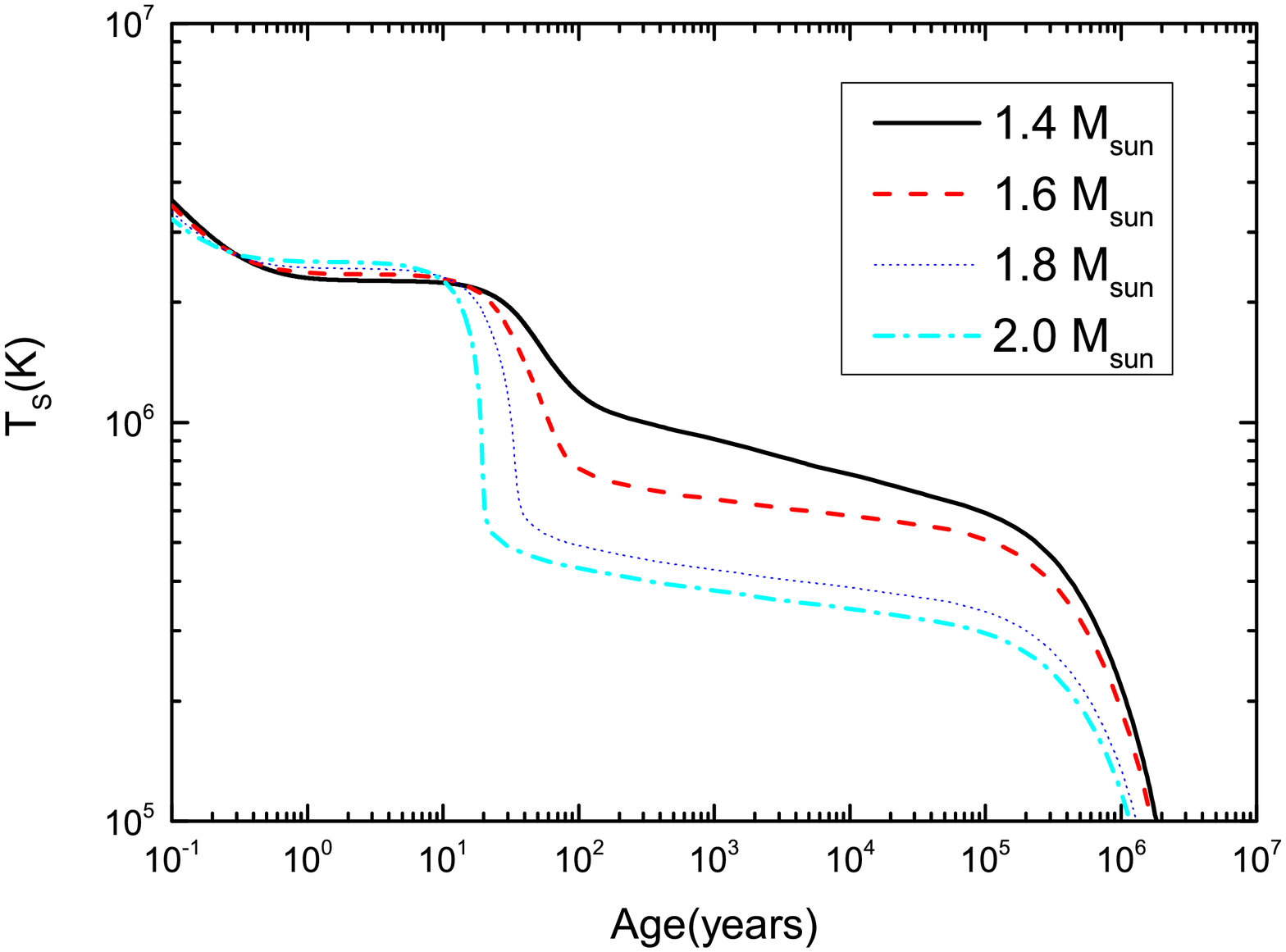}
\caption{Surface temperature evolution for neutron stars under the QMC model (with a pasta phase) and nucleon pairing. \label{coolQMCpasta1SF}}
\end{figure}

We begin by showing the thermal evolution of the QMC model (with pasta), which is shown in Fig.~\ref{coolQMCpasta1SF}.

\begin{figure}[!htb]
\includegraphics[scale=0.34]{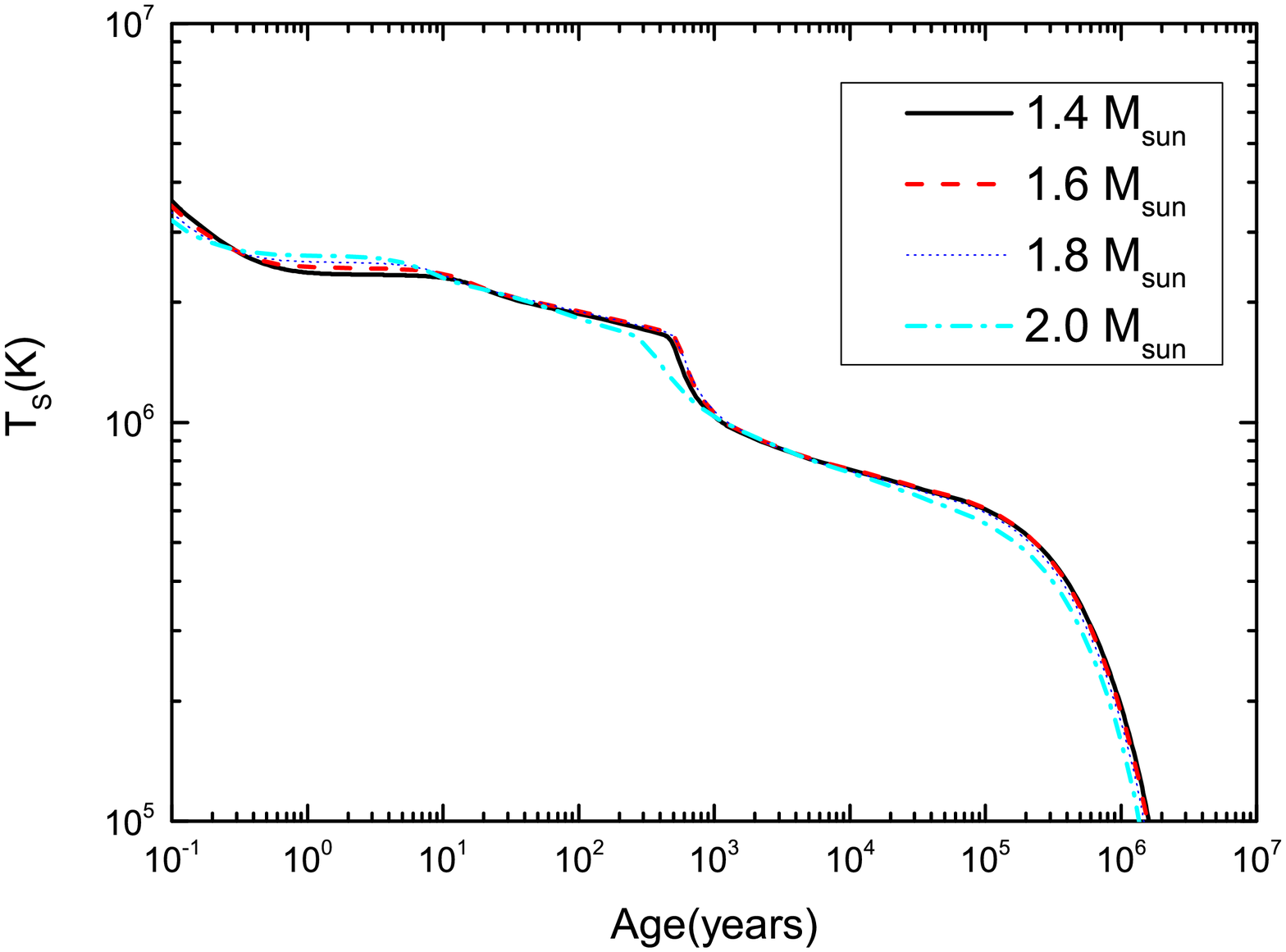}
\caption{Same as in Fig.~\ref{coolQMCpasta1SF} but for the QMC$\omega \rho$ model (with pasta). \label{coolQMCwrpasta1SF}}
\end{figure}

Fig.~\ref{coolQMCwrpasta1SF} shows the surface temperature evolution for the stars under the QMC$\omega \rho$ model with nucleonic pairing.

The results illustrated in Figs.~\ref{coolQMCpasta1SF} and \ref{coolQMCwrpasta1SF} show us that, as expected, the presence of nucleon pairing leads to slower cooling (as neutrino emission processes involving paired nucleons are exponentially suppressed, once the superfluid phase is achieved). For the QMC$\omega \rho$ this is less relevant, as in this case the DU process was already suppressed due to the nature of this equation of state itself, which  is connected to lower proton fractions. For this model, however, we see that a "second knee" appears near the 1000 year age. This second temperature drop is associated with the onset of neutron supefluidity on the outer layers of the star, which gives rise to a burst of neutrino emissions (associated with the Pair Breaking-formation process), leading to a sudden temperature drop. This phenomena has been used in previous studies to explain the apparent temperature drop of the neutron star in Cas A \cite{Page2011a,Shternin2011}, although Cas A  itself has been put into question recently \cite{Posselt2018}. 

The results of the QMC (pasta) model (fig.~\ref{coolQMCpasta1SF}) are more interesting for the purposes of this research. We see that with the presence of pairing, the cooling exhibits a "broader spectrum", with higher masses associated with faster cooling, smaller masses with lower, and intermediate masses in between. Such behavior makes this model ideal to confront against observational data, as it is more likely to fall into the range of observed neutron star temperatures. 

\begin{figure}[!htb]
\includegraphics[scale=0.34]{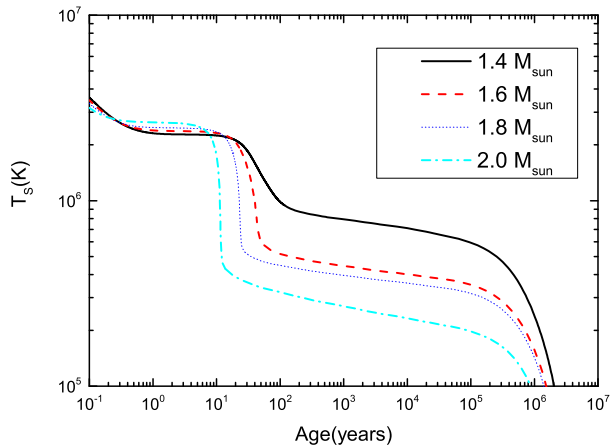}
\caption{Surface temperature evolution for neutron stars under the MQMC1  model and nucleon pairing. \label{coolQMC1SF}}
\end{figure}

\begin{figure}[!htb]
\includegraphics[scale=0.34]{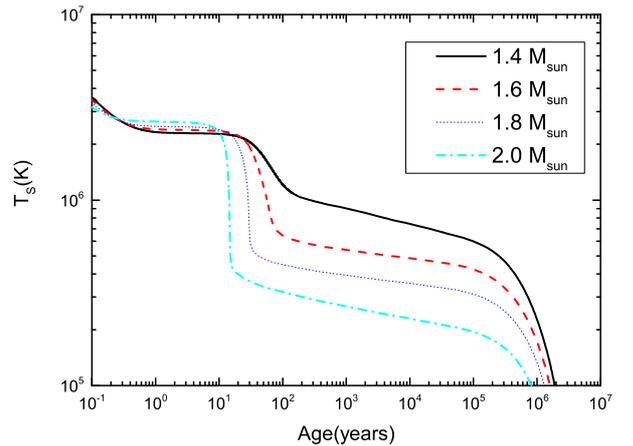}
\caption{Surface temperature evolution for neutron stars under the MQMC2  model and nucleon pairing. \label{coolQMC2SF}}
\end{figure}

We have also considered the thermal evolution of the MQMC stars with superfluidity. In here we show only the cooling of the stars of the MQMC1 and MQMC2, which are shown in Figs.~\ref{coolQMC1SF} and ~\ref{coolQMC2SF}. The cooling of superfluid stars of the MQMC3 model is omitted as they exhibit a similar behavior of that shown in Fig.~\ref{coolQMCpasta1SF}.

\begin{figure}[!htb]
\includegraphics[scale=0.34]{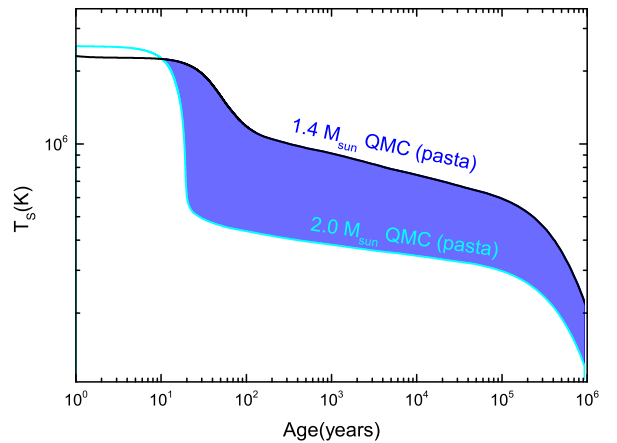}
\caption{Thermal evolution of neutron stars under the QMC model, with pasta phase and nucleon superfluidity. Blue shaded region represents the possible thermal evolution for stars with different masses under this model \label{QMCband}}
\end{figure}

\begin{figure}[!htb]
\includegraphics[scale=0.34]{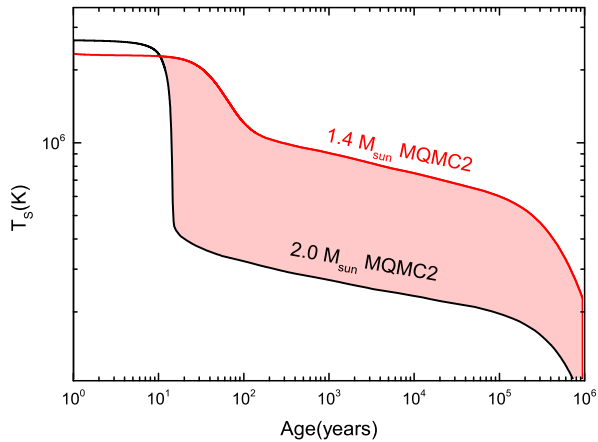}
\caption{Thermal evolution of neutron stars under the MQMC2 model with nucleon superfluidity. Pink shaded region represents the possible thermal evolution for stars with different masses under this model \label{MQMCband}}
\end{figure}

These results show us that both MQMC1 and MQMC2 models lead to satisfactory thermal evolution behavior, leading to a large band covered by different stars masses that could potentially describe the observed data. As it is usual in thermal evolution studies, one cannot (usually) pin point the composition of the star by such investigations but rather rule out models that are less likely to describe observed data. We now compare the band of possible thermal evolutions (spanned by low-to-high mass stars) in the two models we have deemed best (in the cooling context) -- these are shown in Figs~\ref{QMCband} and \ref{MQMCband}.

\begin{figure}[!htb]
\includegraphics[scale=0.34]{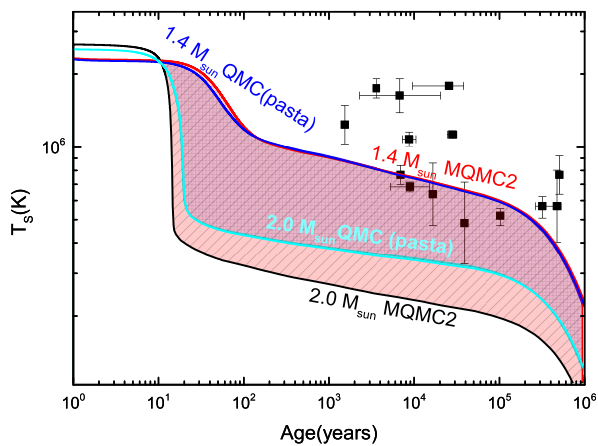}
\caption{Two best thermal models and their respective cooling band against a set of prominent observed neutron stars thermal properties.  \label{2modelband}}
\end{figure}

\begin{figure}[!htb]
\includegraphics[scale=0.34]{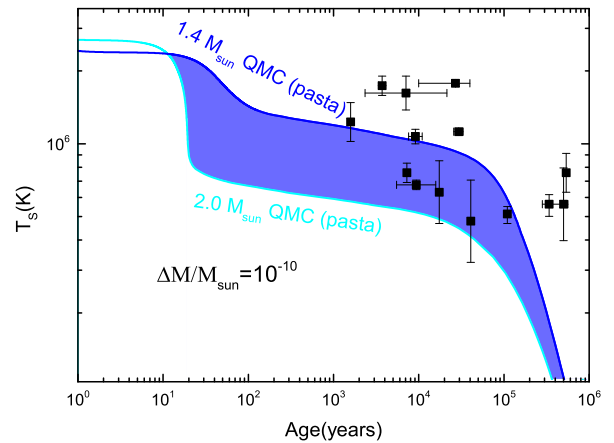}
\caption{Thermal evolution of neutron stars of the QMC (with pasta) model, and an accreted atmosphere of $\Delta M/M = 10^{-10}$.  \label{DMcool}}
\end{figure}

We also confront the results of Figs~\ref{QMCband} and \ref{MQMCband} with most prominent thermal observed data \citep{Beloin2018,SafiHarb2008,Zavlin1999,Pavlov2002,Mereghetti1996,Zavlin2007,Pavlov2001,Gotthelf2002,McGowan2004,Klochkov2015,McGowan2003,McGowan2006,Possenti1996,Halpern1997,Pons2002,Burwitz2003,Kaplan2003,Zavlin,Ho2015}.
 -- as shown in Fig.~\ref{2modelband}, from where we see that 
the major difference between the two models is that the MQMC2 spans a wider band, covering more of the low temperature region. We do not however have any data available on such region, possibly due to instrument sensitivity limitations. What we can see, however, is that both models have difficulties in matching the higher temperatures observed. This could be due to the presence of some unaccounted heating process or even to the uncertainties on the estimation of the star age. This can be remedied somewhat by taking into account a more sophisticated model for the neutron star atmosphere. For the present model we have only considered standard neutron star atmosphere \cite{Gudmundsson1983}. At the neutron star formation, some accretion might take place \cite{Potekhin1997}, depositing some materials onto the surface of the star. In order to account for that we consider the QMC model (with pasta and superfluidity) and allow for an accreted atmosphere of $\Delta M = 10^{-10} M$. The results are shown in Fig.~\ref{DMcool} and they show that the accreted atmosphere allow the surface to reach higher temperatures, making the agreement with observed data a little better. There is price paid, however, that is the speed up of cooling at later ages - this can be explained due to higher temperatures achieved by the star surface, that also enhance black-body photon emission (which is the dominant cooling mechanism in the late stages of evolution). We also see that a modest accreted atmosphere is not enough to explain the higher temperature stars -- which is in agreement with other studies (see \cite{Negreiros2010} for instance). One must also keep in mind that there is great uncertainty regarding the age of the star \cite{Page2004}. 

\section{Final Remarks}
\label{final}

In the present work we have used five different versions of the QMC model to compute the Love number and tidal polarizability and confront them with GW170817 constraints and also to investigate the cooling process they describe. Two of the models are based on the original bag potential structure and three versions consider a harmonic oscillator potential to confine the quarks. The bag-like models also incorporate the pasta phase used to describe the inner crust of neutron stars. 

We have compared our EOS with the FSUGarnet \cite{jorge2018} and checked that
the outer crusts are coincident and the liquid core of all EOS are very similar, 
most of the differences residing on the inner crust and around the crust-core transition region, where the pasta phase was included. Our results point to  the fact that the pasta phase plays only a minor role in the calculation of the dimensionless tidal polarizability and just two of the models (QMC$\omega\rho$ and MQMC3)
give results that lie within the expected range of the canonical star $\Lambda_{1.4}$ and, at the same time, appear inside the region delimited by the confidence lines. At this point we see no clear correlation between the model symmetry energy or its slope and the fact that it satisfies (or not) the constraints investigated.

We have then used the very same models to study the cooling process and have verified that, in this case, there is a clear correlation between the slope of the symmetry energy and the velocity of the cooling process, i.e., models with higher (lower) values of the slope produce fast (slow) cooling. The pasta phase, once again, plays a minor role in the cooling process by speeding up the process in about 3.5 years if superconductivity is not considered. The models that seems to better describe the observational data are the QMC (with pasta) and MQMC2, but none can reach the region of very high temperatures. 

In fact, our studies show that we cannot pin down one unique model that can, at the same time, describe astrophysical quantities and a perfect description of the possible cooling processes and more observational data is necessary.

\begin{acknowledgments}

This work is a part of the project INCT-FNA Proc. No. 464898/2014-5, was partially supported by CNPq~(Brazil) under grants 301155.2017-8~(D.P.M.), 310242/2017-7 and 406958/2018-1~(O.L.), 308486/2015-3~(T.F.) and 433369/2018-3~(M.D.), by Capes-PNPD program~(C.V.F.), and by Funda\c{c}\~ao de Amparo \`a Pesquisa do Estado de S\~ao Paulo~(FAPESP) under the thematic projects 2013/26258-4 (O.L., T.F.) and 2017/05660-0~(T.F.). R.N. also acknowledges that this project was partly funded by FAPERJ, under grant E-26/203.299/2017.

\end{acknowledgments}

\end{document}